\documentclass[11pt]{article}
\pdfoutput = 1

\usepackage[utf8]{inputenc}
\usepackage{color,graphicx}
\usepackage{amsmath}
\usepackage{verbatim}
\usepackage{amssymb}
\usepackage{mathabx}
\usepackage{physics}
\usepackage{amsfonts}
\usepackage{cite}
\usepackage{array}
\usepackage{setspace}
\usepackage{float}
\usepackage{url}
\usepackage{mathtools}
\usepackage{bbold}
\usepackage{tikz}
\usepackage{graphicx}
\usepackage{subfigure}
\usepackage[labelfont=bf,font={sf}]{caption}
\usepackage{mathrsfs}
\usepackage{fancybox}
\usepackage{tensor}
\usepackage{tikz}
\usepackage{cite}
\allowdisplaybreaks

\usepackage[margin = 2.2cm]{geometry}
\usepackage[ragged]{footmisc}
\usepackage{hyperref}
\hypersetup{
    colorlinks,
    citecolor=blue,
    filecolor=black,
    linkcolor=blue,
    urlcolor=black,
    linktocpage=true
}

\def\nn{\nonumber}

\newcommand{\be}{\begin{equation}}
\newcommand{\ee}{\end{equation}}
\newcommand{\bea}{\begin{align}}
\newcommand{\eea}{\end{align}}
\newcommand{\bi}{\begin{itemize}}
\newcommand{\ei}{\end{itemize}}

\newcommand{\jt}{{\mathrm{JT}}}

\def\t{\tau}

\def\r{\rho}

\numberwithin{equation}{section}

\title{Template}

\begin{document}

\thispagestyle{empty}
\begin{center}
\vspace*{.4cm}
     {\LARGE \bf 
  Comments on Lorentzian topology change in JT gravity
  }
    
    \vspace{0.4in}
    {\bf Mykhaylo Usatyuk}

    \vspace{0.4in}
{Center for Theoretical Physics and Department of Physics, University of California, Berkeley,\newline{} CA 94720, USA}
    \vspace{0.1in}
    
    {\tt musatyuk@berkeley.edu}
\end{center}

\vspace{0.4in}

\begin{abstract}
\noindent We propose a definition for the Lorentzian Jackiw-Teitelboim (JT) gravity path integral that includes Lorentzian topology changing configurations. The construction is inspired by the bosonic string genus expansion on singular Lorentzian worldsheets, with geometries known as lightcone diagrams playing a prominent role. The Lorentzian path integral is defined through a suitable analytic continuation of the Euclidean path integral, and includes metrics that become degenerate at isolated points allowing for Lorentzian topology changing transitions. We discuss the relation between Euclidean JT amplitudes and the proposed Lorentzian amplitudes.

\end{abstract}

\pagebreak
\setcounter{page}{1}
\tableofcontents
\section{Introduction}
Euclidean wormholes have played an important role in recent developments in quantum gravity. The study of two dimensional Jackiw-Teitelboim (JT) gravity \cite{Saad:2019lba} has been central to these developments. Topology changing Euclidean configurations have been found to be important in the calculation of a unitary page curve \cite{Almheiri:2019qdq,Penington:2019kki}, the spectral form factor\cite{Saad:2018bqo,Saad:2019lba,Blommaert:2022lbh}, and a variety of other interesting effects, see \cite{Saad:2019pqd,Iliesiu:2021ari,Stanford:2022fdt} for a few selected examples.

However, so far a satisfactory Lorentzian explanation for Euclidean wormhole calculations has been lacking. The standard procedure is to calculate a Euclidean amplitude and then analytically continue it to obtain a Lorentzian result. The role that Euclidean wormholes play in the Lorentzian theory is not apparent in this analytic continuation. Similarly, in the standard canonical quantization treatment of Lorentzian JT gravity\cite{Harlow:2018tqv} the topology of the Lorentzian spacetime is fixed, and it's unclear how the Euclidean path integral formulation of the theory\cite{Saad:2019lba} can be related to the Lorentzian formulation. 

\begin{figure}[H]
\includegraphics[width=6cm]{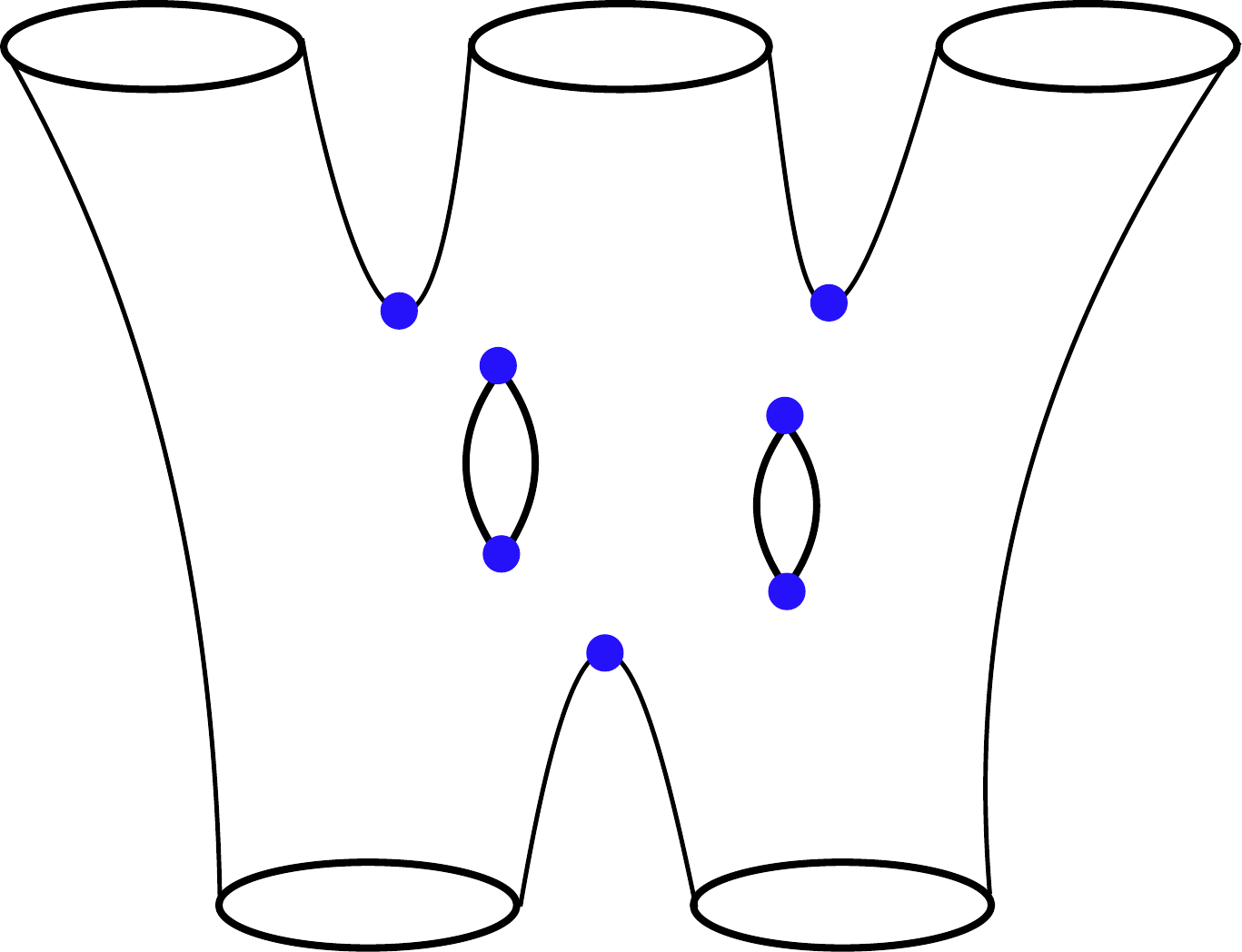}
\centering
\caption{Lorentzian topology changing transition where two spatial circles evolves into three circles. The metric is Lorentzian everywhere except the splitting points (blue circles) where it is degenerate. At the splitting points the topology of spatial slices changes. The spacetime is an analytic continuation of a genus two Euclidean geometry with five circular boundaries.}
\label{fig:example}
\end{figure}

In this work we formulate a Lorentzian theory of JT gravity that includes Lorentzian topology changing configurations. The theory is defined through a special analytic continuation of the standard Euclidean path integral. Euclidean wormholes are turned into topology changing Lorentzian configurations with degenerate points in the metric, see figure \ref{fig:example}.
Our proposal is inspired by a formulation of bosonic string theory on degenerate Lorentzian worldsheets known as the interacting string picture\cite{Mandelstam:1973jk,Mandelstam:1985ww}. In \cite{Wol86,GidHok87} it was argued that the interacting string picture is equivalent to the standard Euclidean path integral formulation to all orders in the genus expansion. This is accomplished by gauge-fixing the Euclidean path integral to lightcone diagrams\cite{Wol86}, which are a special class of metrics that can be analytically continued to Lorentzian signature where they become the topology changing geometries of the interacting string picture.  

To define the Lorentzian JT path integral we closely follow the construction of the bosonic string genus expansion with singular Euclidean/Lorentzian worldsheets \cite{Wol86,GidHok87,Sonoda:1987ra}. We start with the standard Euclidean JT path integral and with a suitable gauge choice and analytic continuation we end up with a Lorentzian path integral over degenerate metrics. We now briefly explain this construction, leaving technical details to the main sections.

\subsection*{Summary of results}
We begin with the two dimensional Euclidean gravity path integral with specified boundary conditions. In this paper we consider boundary conditions given by $n$ geodesic circles of given lengths $\Vec{b}=(b_1, \cdots, b_n)$. For simplicity we assume the bulk geometry has fixed genus $g$ and is fully connected. The path integral is computed by
\be
Z = \int \frac{\mathcal{D}g_{\mu \nu} \mathcal{D}\Phi}{\text{Vol}} e^{-I_{\jt}[g,\Phi]} =\hspace{.7mm}\raisebox{-18mm}{\includegraphics[width=3.5cm]{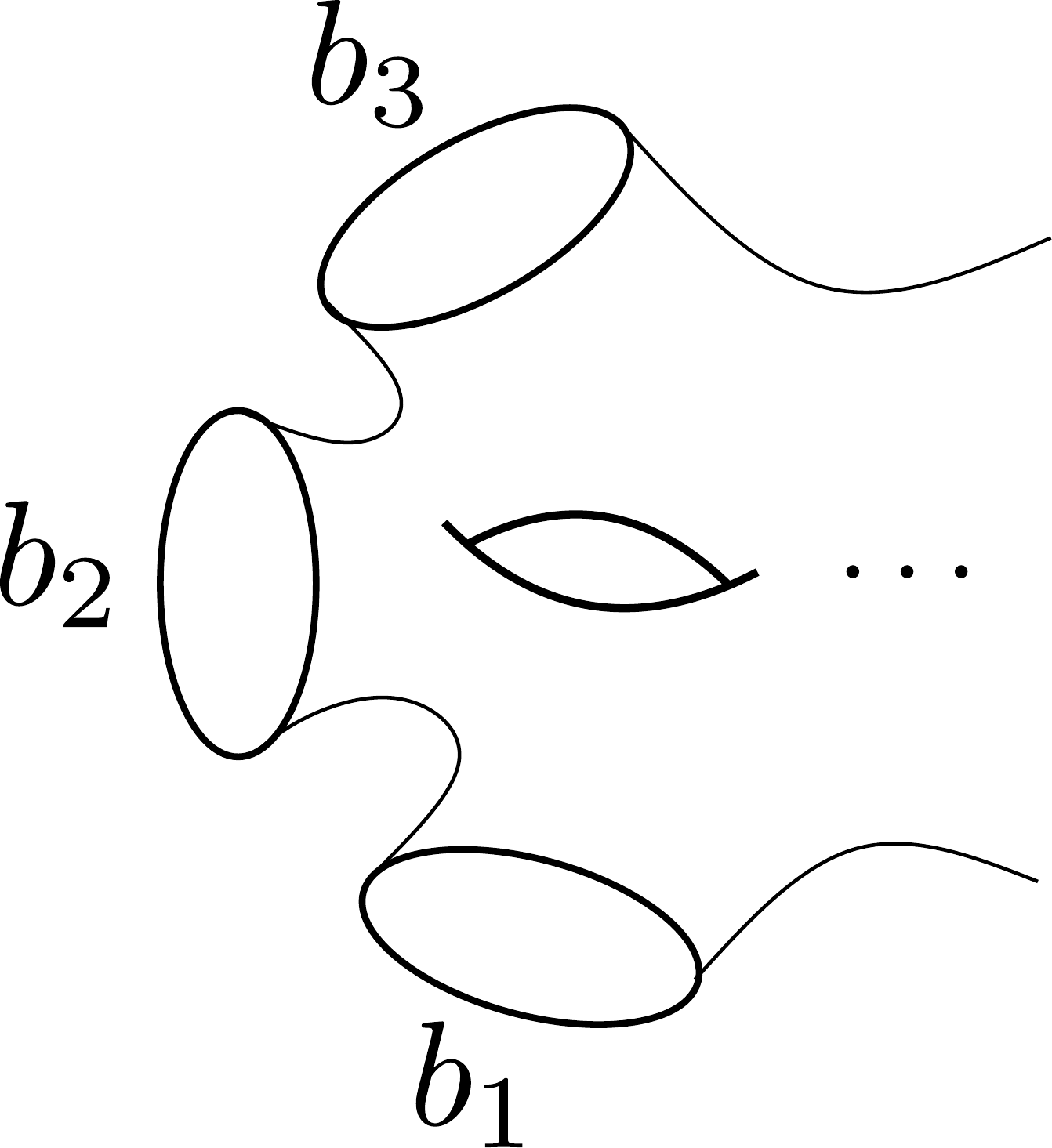}}
\ee
The integral over Euclidean metrics can be split into an integral over the Weyl factor $\omega$ and over the moduli $\hat{g}$, where all metrics can be represented as $g = e^{2\omega} \hat{g}$. The moduli space of $\hat{g}$ is the moduli space of punctured Riemann surfaces $\mathcal{M}_{g, n}$ of genus $g$ with $n$ punctures. It was shown by Giddings and Wolpert \cite{Wol86} that this Moduli space has a representative metric $\hat{g}$ that is flat with curvature singularities at isolated points, this is known as a Euclidean lightcone metric. Choosing the Euclidean lightcone metric as our representative metric $\hat{g}$ gives us a path integral over singular Euclidean geometries
\be
Z =\int_{\rm moduli} {\mathrm d} ({\rm measure}) \int \mathcal{D} \omega \mathcal{D}\Phi e^{-I_{\jt}[e^{2\omega} \hat{g},\Phi]}=\hspace{.0mm}\raisebox{-18mm}{\includegraphics[width=2.8cm]{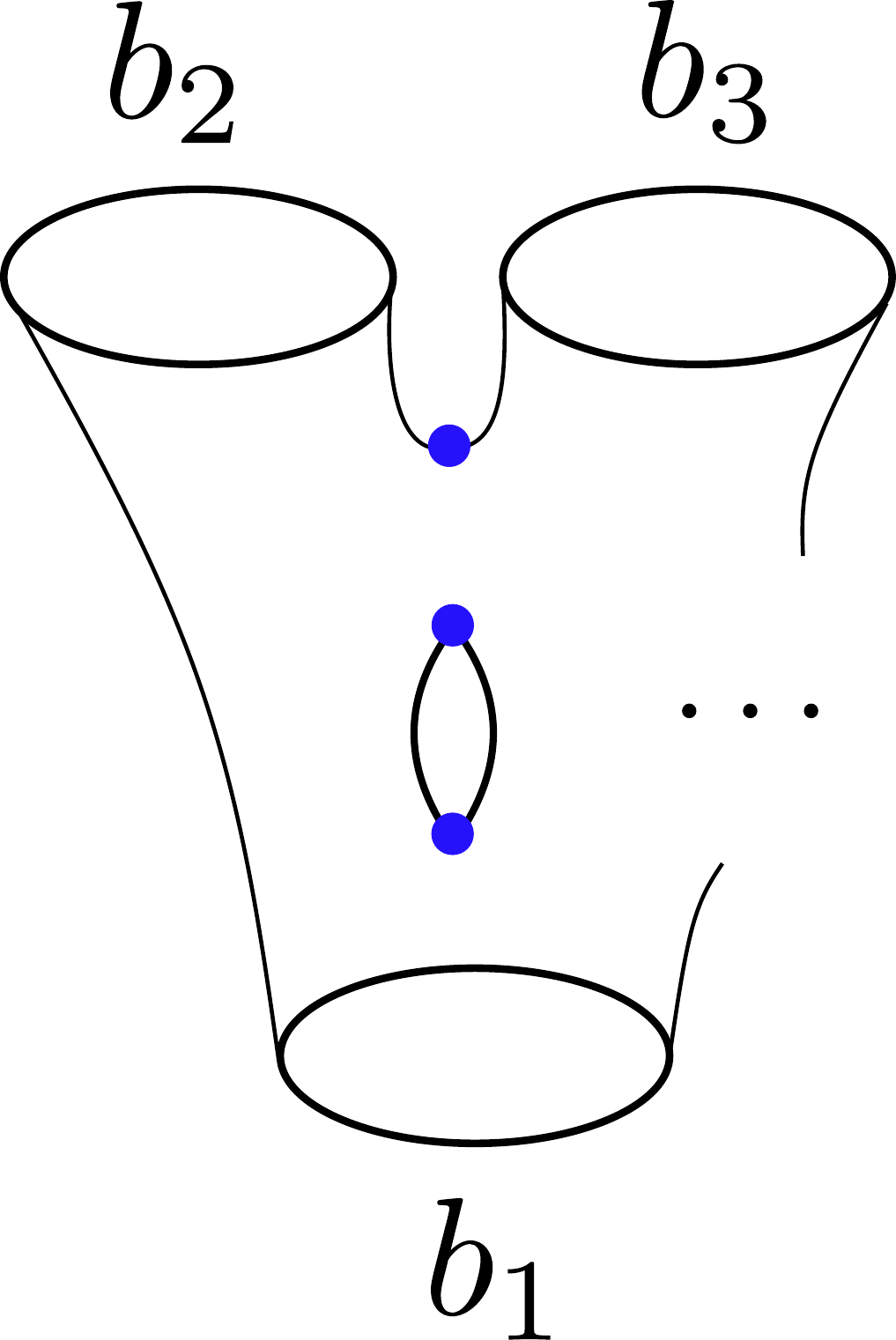}}
\ee
In the above figure, the circles (blue) corresponds to points where the metric $\hat{g}$ becomes degenerate $\det \hat{g} = 0$. It is at these points that the topology of spatial slices changes. The integral over the moduli is partially over all possible locations of the degenerate points. The Euclidean lightcone diagram has a globally defined euclidean time $\tau$ in terms of which the metric is flat everywhere except the degenerate points. The boundary conditions are now specified at  $\tau=\pm \infty$, and we must make a choice to send boundary conditions either to the future or the past. In the above figure the boundary of length $b_1$ has been sent to the past while the boundaries of length $b_2, b_3$ have been sent to the future.

To turn the above Euclidean path integral into a Lorentzian path integral, we will analytically continue the Euclidean lightcone geometries. Since the time $\tau$ is globally defined, we can analytically continue $\tau \to i \tau$ to get a Lorentzian signature metric $\hat{g}$, known as a Lorentzian lightcone diagram. We now have a path integral over degenerate Lorentzian metrics that incorporate topology changing transitions
\be
Z_{\mathrm{L}} = \int_{\rm moduli} {\mathrm d} ({\rm measure}) \int \mathcal{D} \omega \mathcal{D}\Phi e^{i I_{\jt}[e^{2\omega} \hat{g},\Phi]}.
\ee
Our \emph{\textbf{definition}} for the Lorentzian JT path integral will be the above integral over Lorentzian lightcone diagrams. The role of the Weyl factor $\omega$ will be to give the geometry constant negative curvature away from the degenerate points.

In the above procedure we started with the usual Euclidean path integral and through a gauge choice and a suitable analytic continuation we ended up with a Lorentzian path integral. It might then be expected that the amplitudes computed with this Lorentzian path integral should agree with the corresponding Euclidean amplitudes. Indeed, this is the argument of D'Hoker and Giddings\cite{GidHok87} that the interacting string picture\cite{Mandelstam:1985ww} is equal to the Euclidean path integral to all orders in the genus expansion\footnote{Giddings and D'Hoker \cite{GidHok87} stopped short of analytically continuing the Euclidean lightcone diagrams to Lorentzian signature. Thus they argued that the analytically continued interacting string picture was equivalent to the usual Euclidean path integral.}. In the case of JT gravity there are some additional subtleties that arise due to the degenerate points, and we return to this question in section \ref{sec:3} and in the discussion. We now briefly summarize the rest of the paper. 

In Section \ref{sec:2} we review the basic aspects of Euclidean and Lorentzian lightcone diagrams. We also review the work of Giddings and Wolpert \cite{Wol86} where it was shown that Euclidean lightcone diagrams give a single cover of the moduli space of punctured Riemann surfaces. Lastly, we discuss the problem of finding a Weyl factor to turn a Euclidean/Lorentzian lightcone metric into a constant negative curvature geometry with degenerate points.

In Section \ref{sec:3} we fill in the technical details of the path integral over lightcone diagrams. We explain how boundary conditions are implemented, and we discuss the integration measure over the moduli space of lightcone diagrams. The inclusion of degenerate points introduces certain ambiguities into the path integral, and we discuss how these ambiguities modify the relation of the Euclidean amplitudes to the Lorentzian amplitudes.

In the Appendices we construct the Lorentzian pair of pants with constant negative curvature, and we include additional details on punctured Riemann surfaces and the integration measure.

%%%%%%%%%%%%%%%%%%%%%%%%%%%%%%%%%%%%%%%%%%%%%%%%%%%%%%%%%%%%%%%%%%%%%%%%%%%%%%%%%%%%%%%%%%%%%%%%%%%%%%%%%%%%%%%%%%%%%%%%%%%%%%%%%%%%%%%%%%%%%%%%%%%%%%%%%%%%%
\section{Lightcone diagrams} \label{sec:2}
In this section we review aspects of Euclidean and Lorentzian lightcone diagrams, their connection to punctured Riemann surfaces, and how to construct constant negative curvature analogues of lightcone diagrams.

\subsection{Euclidean and Lorentzian lightcone diagrams}
\begin{figure}
\includegraphics[width=11cm]{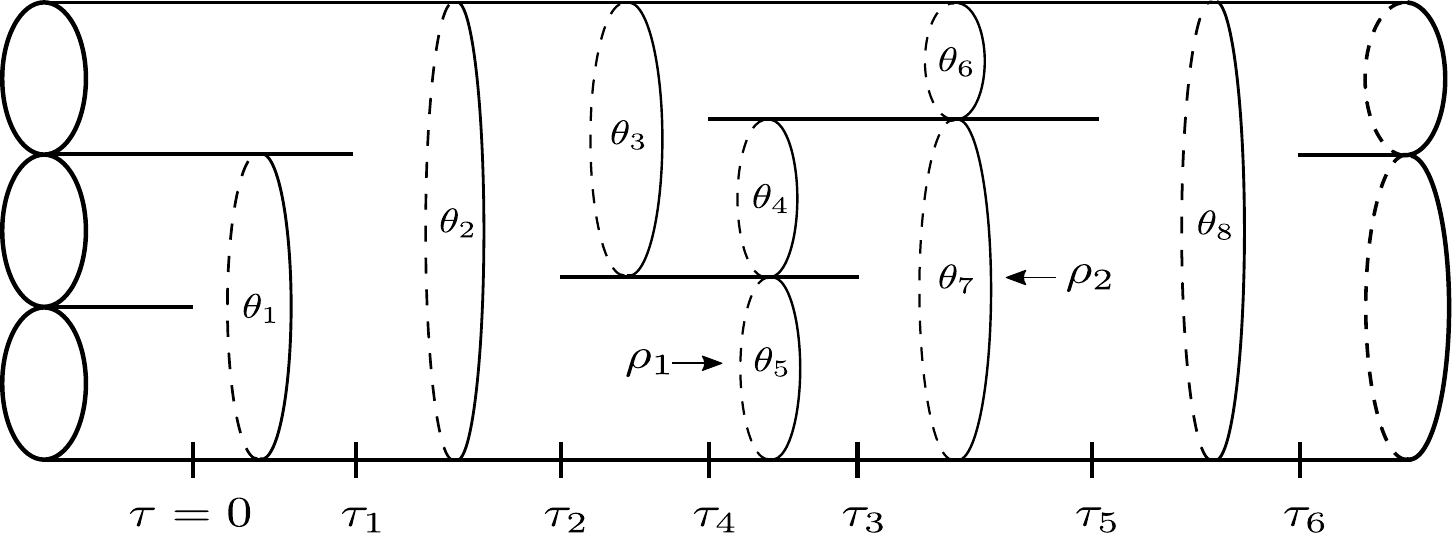}
\centering
\caption{Euclidean/Lorentzian lightcone diagram of genus two with five boundaries. The geometry is flat except at interactions points $\tau_i$ with delta functions of curvature. A lightcone diagram is built by gluing pairs of pants together. The interaction times are labelled by $\tau$, the twist angles by $\theta$, and the free internal radius by $\r$. }
\label{fig:ELC}
\end{figure}

\begin{figure}
\includegraphics[width=10cm]{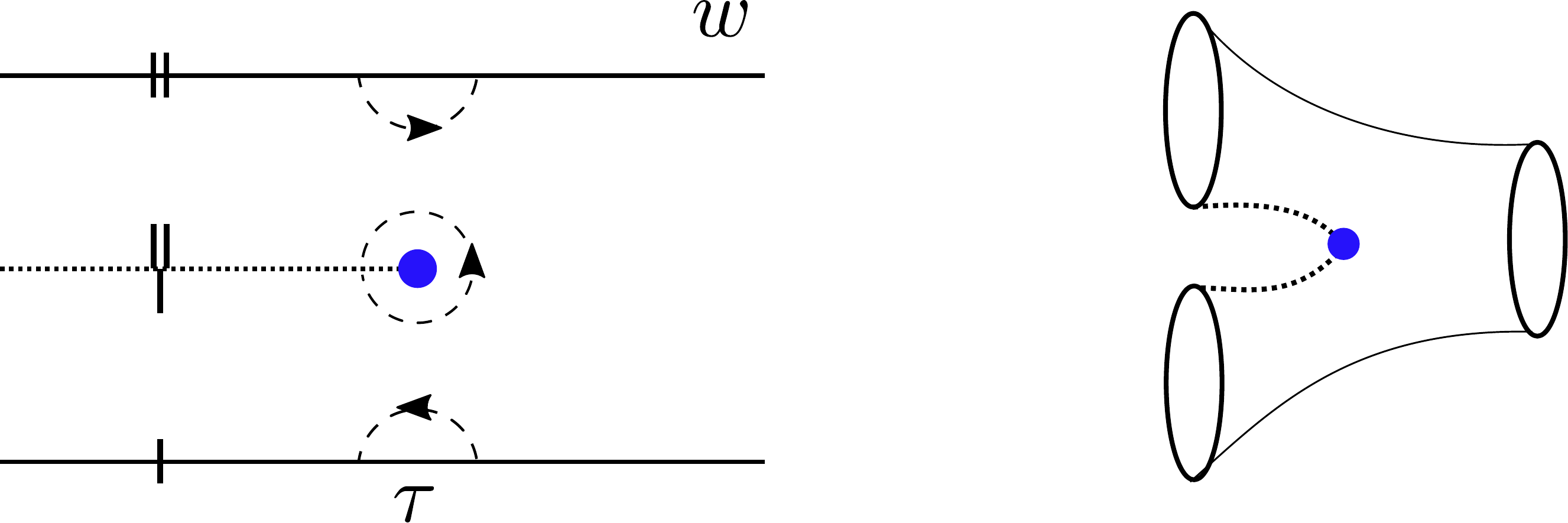}
\centering
\caption{Euclidean Pair of pants constructed by identification in the complex $w=\tau + i \sigma$ plane. The middle line (dashed) is infinitesimally split up to the interaction time $\tau$ and identified with the top and bottom segments as indicated. The singular point is indicated by the circle (blue). A closed loop around the singular point goes through angle $4\pi$.}
\label{fig:wcoords}
\end{figure}

A Euclidean/Lorentzian lightcone diagram is a two dimensional geometry with degenerate metric $g$ built out of flat Euclidean/Lorentzian cylinders joined together at singular points, see figure \ref{fig:ELC}. The two basic properties of a lightcone diagram are the number of asymptotic cylinders $n$ and the genus $g$. The number of cylinders running off to infinity is given by $n \geq 2$ with a fixed number extending to past or future infinity\footnote{At least one cylinder must run to the future and one to the past.}. In the next section we will see that in/out states are specified by introducing boundary conditions on the asymptotic cylinders. 
The genus $g$ and number of boundaries $n$ determine the number of times the diagram splits apart and joins together in the interior of the geometry. At the splitting and joining points the metric is degenerate, and there are $2g-2+n$ such degenerate points. We will also call these interaction/singular points, and denote where they occur by the subscript $z_I$. The curvature on a Euclidean lightcone diagram is given by
\be \label{eqn:metric_ELC}
\frac{1}{2}\sqrt{g} R= - 2 \pi \sum_{I=1}^{2g-2+n} \delta^2(z-z_I), \qquad \qquad  \det g(z_I)=0.
\ee
Note that since the determinant of the metric is zero at the interaction points the action on a lightcone diagram can be badly behaved at these points. Even though the metric is singular, the singularity is sufficiently mild. The Gauss-Bonnet theorem continues to hold
\be
\frac{1}{2}\int \sqrt{g} R = 2\pi \chi,
\ee
where $\chi = 2-2g-n$. All the curvature required for Gauss-Bonnet to be satisfied is localized at the interaction points. In Euclidean signature we introduce coordinates $\tau, \sigma$ on the diagrams where $\tau$ is a time coordinate and $\sigma$ is a periodic coordinate around the cylinder. We can represent the diagrams by making identifications in the complex plane $w=\tau + i \sigma$, see Fig. \ref{fig:wcoords} for an example of how to build a pair of pants. To travel around a singular point we must go through angle $4\pi$ in $w$ coordinates, so the singularity is a double cover of the Euclidean plane. 

We can easily find the metric near the singular points using complex coordinates
\be \label{eqn:ELC_metric}
ds^2 = |z|^{2\alpha} dz d \bar{z} = e^{2\omega} dz d \bar{z}, \qquad \qquad \frac{1}{2}\sqrt{g} R = -2\pi \alpha \delta^{(2)}(z).
\ee
which is a cone of opening angle $2\pi(1+\alpha)$ with conformal mode $\omega = \frac{1}{2}\log |z|^{2\alpha}$. The singularities of the Euclidean lightcone diagram correspond to $\alpha=1$. We can see this by the coordinate $w=z^2$, going through angle $2\pi$ in $z$ takes us through an angle of $4\pi$ in $w$, which is a double cover of the Euclidean plane as required.

\subsection*{Lorentzian lightcone diagrams}
The above discussion was largely restricted to Euclidean lightcone diagrams. However, by an analytic continuation of the time coordinate $\tau \to i \tau$ we obtain a metric that is Lorentzian everywhere except at the degenerate points. These geometries are Lorentzian lightcone diagrams, and the metrics on them are known as almost Lorentzian metrics. The curvature is given by
\be 
\frac{1}{2}\sqrt{-g} R = 2 \pi i \sum_{I=1}^{2g-2+n}\delta^2(x-x_I), \qquad \qquad  \det g(x_I) =0. \label{eqn:metric_LLC}
\ee
These geometries are constructed by gluing Lorentzian pants along geodesic boundaries. The pair of pants cannot be given an everywhere Lorentzian metric, but with the inclusion of the above singularity at the splitting point it becomes possible to equip it with an almost Lorentzian metric \cite{Louko:1995jw,Witten:2021nzp}. We can explicitly write down a regularized complex metric near the splitting point of the pants. In the conventions of \cite{Witten:2021nzp} this metric is given by
\be
ds^2 = \left(x^{2}+y^{2}+ \gamma \right)\left(d x^{2}+d y^{2}\right)-(2 \pm i \epsilon)(x d x - y d y)^{2}.
\ee
The splitting point is located at $x=y=0$, and we have introduced regulators $\gamma, \epsilon > 0$. The almost Lorentzian pants are obtained from the above metric as we take the regulators to zero, with appropriate identifications of the $x,y$ plane\cite{Louko:1995jw}. Analogous to the Euclidean case, the almost Lorentzian pants are a double cover of Minkowski space. The $\gamma$ regulator temporarily removes the degenerate point and the $\epsilon$ regulator deforms the metric into what is known as an allowable complex metric\footnote{A complex metric is defined as allowable if the following condition is satisfied\cite{Witten:2021nzp,Kontsevich:2021dmb}. Consider a basis where the metric is diagonal $g_{i j} = \lambda_i \delta_{i j}$, if $\sum_i |\operatorname{Arg} \lambda_i| < \pi$ at every point on the manifold then the metric is allowable. Allowable metrics satisfy certain nice properties. For example, this condition ensures that the path integral of a $p$-form gauge field converges on the background $g$.} \cite{Witten:2021nzp}. This is important for us because allowable complex metrics satisfy an analytically continued version of the Gauss-Bonnet theorem \cite{Louko:1995jw,Witten:2021nzp}. 

Ultimately we will be integrating over all lightcone diagrams, and our Euclidean action will contain a topological term suppressing each geometry by it's Euler characteristic $e^{S_0 \chi}$, where $S_0$ is some large constant. Upon analytic continuation of the time coordinate the topological term will be proportional to
\be
\frac{i}{2} \int d^2 x \sqrt{-g} R. 
\ee
For allowable complex metrics the above integral is equal to an analytically continued version of the Gauss-Bonnet theorem \cite{Witten:2021nzp}
\be \label{eqn:L_GaussBonnet}
\frac{1}{2} \int d^{2} x \sqrt{-g} R=-2 \pi i \chi.
\ee
All metrics considered in this paper can be deformed to be allowable, and so we maintain the same topological expansion in Lorentzian as in Euclidean signature. We can directly see that the above formula holds for the Lorentzian lightcone metric \eqref{eqn:metric_LLC}. An important property is that given an allowable complex metric $g$, $e^{2\omega} g$ is also allowable if $\omega$ is real. Since the Lorentzian pants are allowable so is any rescaling of the pants, which will be useful when we need to consider the constant negative curvature pants geometry.

\subsection*{Moduli space of lightcone diagrams}
We now explain the moduli space of inequivalent Euclidean/Lorentzian lightcone diagrams of genus $g$ with $n$ boundaries. The starting point is to specify the geodesic radius $r_i$ of each of the $n$ boundaries and whether they run off to past or future infinity. A positive radius indicates that the boundary originates from past infinity, a negative radius indicates the boundary goes to future infinity. We demand that the sum of the lengths vanishes $\sum_i r_i = 0$. So that the total length of incoming boundaries is equal to the total length of outgoing boundaries. We denote the total length of incoming/outgoing boundaries by $r_{\rm max}$. 

A general diagram is constructed by gluing Euclidean/Lorentzian pairs of pants together along geodesic boundaries\footnote{The pair of pants used in the construction have boundaries with waist of length $b_1$ and legs of lengths $b_2, b_3$ with the constraint that $b_1 = b_2 + b_3$. There are other Lorentzian pair of pants geometries without this constraint, but they are not used in this construction.}. The different ways to glue pants together gives the moduli space of lightcone diagrams. From figure \ref{fig:ELC}, the moduli space is parameterized by the following coordinates
\begin{align} \nn
    &\text{Interaction Times }  &&\tau_{i} \in [0,\infty)  \qquad \qquad ~\hspace{.6mm} i= 1,  \ldots, 2g+n-3  \\ \nn
    &\text{Twist Angles } &&\theta_{j} \in [0, 2\pi) \qquad \qquad j = 1, \ldots, 3g+n-3 \\ \nn
    &\text{Internal Radius } &&\r_{k} \in [0, r_{\rm max}] \qquad ~~~ \hspace{.1mm} k = 1,\ldots, g
\end{align}
The moduli space is of real dimension $6g-6+2n$. The interaction times $\tau$ define the points at which the cylinders split apart at degenerate points. The diagram can be translated in time so the first interaction time can always be set at $\tau=0$. When gluing two cylinders together we can twist them by a relative angle $\theta \in [0,2\pi)$, this gives the twist angles. Since the pair of pants conserves total geodesic boundary length, the sum of radii of the diagram at any intermediate time must be identical to the sum of beginning or ending radii. We have a freedom in how the total geodesic length is distributed at intermediate times between cylinders, which gives the last parameter $\r_k$ where we integrate over lengths of intermediate cylinders subject to the constraint that the total length is fixed to $r_{\rm max}$.

The end result is that the integral over all lightcone diagrams is given by integrating over interaction times $\tau$, twist angles $\theta$, and intermediate radii $\r$
\be \label{eqn:int_range}
\int[d \tau][d \theta][\r d \r] \equiv \frac{1}{S} \left( \prod_{i=1}^{2g+n-3}  \int_{0}^{\infty} d\tau_i \right) \left( \prod_{j=1}^{3g+n-3} \int_{0}^{2\pi} d \theta_j \right) \left( \prod_{k=1}^{g} \int_0^{r_{\rm max}} \r_k d \r_k \right).
\ee
The above integration range actually overcounts identical lightcone diagrams, and we must divide by a symmetry factor $S$ to integrate over one copy of the moduli space\cite{Wol86}. This symmetry factor has never been explicitly computed, but see \cite{Freidel:2014aqa} for a recent discussion.

\subsection*{Punctured Riemann surfaces and lightcone diagrams}
In this section we'd like to explain the connection between punctured Riemann surfaces and lightcone diagrams. Giddings and Wolpert\cite{Wol86} showed that the moduli space of Euclidean lightcone diagrams gives a single cover of the moduli space of punctured Riemann surfaces $\mathcal{M}_{g,n}$. This is accomplished by showing that every Riemann surface with $n\geq 2$ punctures can be equipped with a unique Euclidean lightcone metric and vice versa. The integral over the moduli space of punctured Riemann surfaces $\mathcal{M}_{g,n}$ can therefore be represented as an integral over the moduli space of Euclidean lightcone diagrams.

We now briefly explain how a given Riemann surface can be equipped with a Euclidean lightcone metric. Consider a genus $g$ Riemann surface $\Sigma$ with local coordinates $z$ and $n$ punctures located at points $z_i$. The goal will be to find a unique meromorphic one-form $\omega = \omega(z) dz$ on $\Sigma$ with simple poles at the punctures $z_i$ with real residues $r_i$ such that their sum is zero $\sum_{i=1}^n r_i = 0$. The specific choice of the $r_i$ does not matter. We also require that the integral of the form $\omega$ is imaginary along any closed cycle\footnote{Without this condition there are many one-forms $\omega$ with the appropriate simple poles, however they fail to give a global lightcone diagram. One way to understand this condition is that on a lightcone diagram a closed loop starts and ends at the same time $\tau$, so around any closed loop $\Delta \tau + i \Delta \sigma = \oint \omega(z) dz$ must be purely imaginary.} on the surface $\Sigma$
\be
\oint \omega (z) dz \in i \mathbb{R}.
\ee
\begin{figure}
\includegraphics[width=10cm]{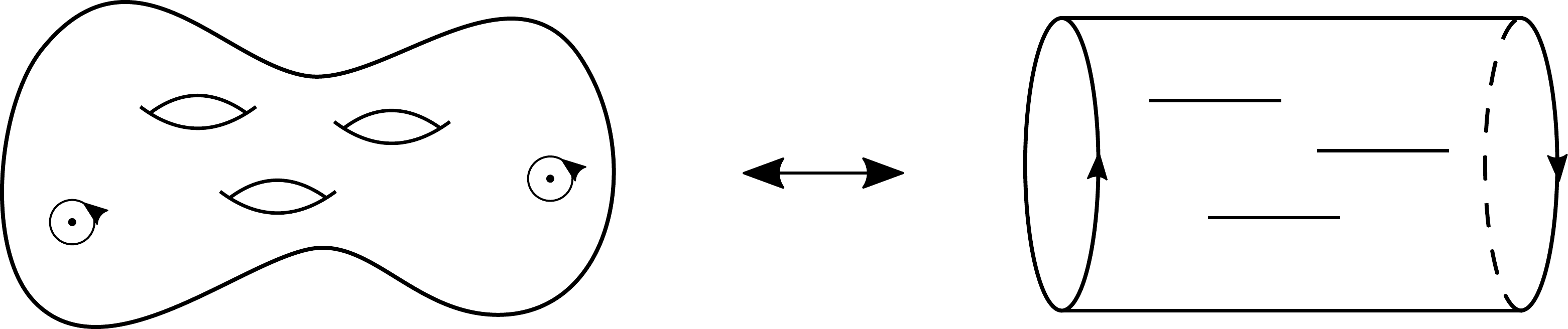}
\centering
\caption{A genus three surface $\Sigma$ with two punctures. We can put a metric on $\Sigma$ given by \eqref{eqn:Wolpert_Metric} to turn it into a Euclidean lightcone diagram with two asymptotic cylinders.}
\label{fig:T3}
\end{figure}
\noindent Under these conditions, Giddings and Wolpert \cite{Wol86} showed that the one-form $\omega$ exists, is unique, and that it defines a unique Euclidean lightcone metric\footnote{The construction of $\omega$ involves the period matrix of the Riemann surface, which is why each Riemann surface is mapped to a unique lightcone diagram.} on $\Sigma$ by 
\be \label{eqn:Wolpert_Metric}
ds^2 = \omega \overline{\omega} = |\omega(z)|^2 d z d \bar{z} = dw d\bar{w},
\ee
see figure \ref{fig:T3}. In the last equality we have defined the lightcone coordinates $w = \tau + i \sigma = \int_{z_0}^z \omega(z) dz$, where $z_0$ is an arbitrary point on $\Sigma$. In these coordinates the metric is explicitly flat with the only breakdown occurring near the poles or zeroes of the form. Near a puncture $z\sim z_i$ the form behaves as $\omega = \frac{r_i}{z-z_i} dz + \ldots$, so the metric will behave as 
\be
ds^2 = \frac{r^2_i}{|z-z_i|^2} dz d\bar{z} + \ldots
\ee
The form will also have $2g-2+n$ zeros at certain points $z_I$ on $\Sigma$, near such points the form behaves as $\omega = \left( z-z_I \right) dz + \ldots$, and the metric will behave as
\be
ds^2 = |z-z_I|^2 dz d\bar{z} + \ldots
\ee
Note that near $z_I$ the metric is that of the interaction point of a Euclidean lightcone diagram in equation \eqref{eqn:ELC_metric}. Thus, the zeros of $\omega$ correspond to degenerate points where the topology of spatial slices of $\Sigma$ change, while the simple poles are located at the punctures.

\begin{figure}
\includegraphics[width=14cm]{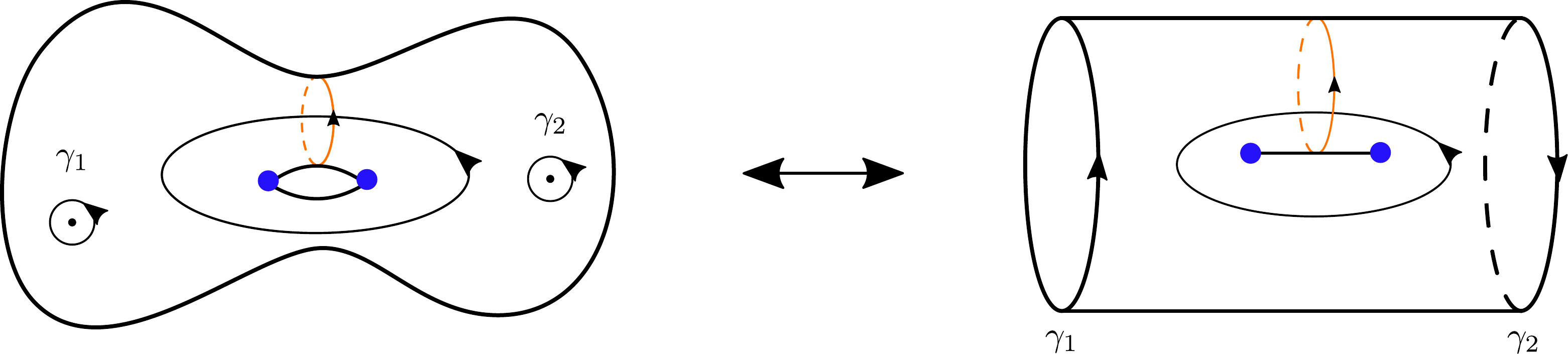}
\centering
\caption{A torus with two punctures mapped to a lightcone diagram with two interaction points and two asymptotic cylinders. The circles (blue) on the torus denote where the one-form $\omega$ has zeros. The two cycles on the torus are illustrated with one at constant time $\tau$ (orange). The loops $\gamma_i$ enclose the punctures which run off to infinity in the lightcone metric. }
\label{fig:torus}
\end{figure}

We now explain how to extract the moduli of the lightcone diagram $(\tau, \theta, \r)$ from the metric \eqref{eqn:Wolpert_Metric}, see figure \ref{fig:torus}. One useful property is that the time coordinate $\tau = \operatorname{Re} \int_{z_0}^z \omega(z) dz$ is globally defined and path independent, since it is defined as the integral of a globally defined one-form\footnote{The integral along two paths can differ by at most an imaginary constant if they enclose a puncture, but the time $\tau$ is the real part of the integral so it does not matter.}. The interaction time differences can be determined by performing an integral between zeros of the form $\tau_{I}-\tau_{J} = \operatorname{Re} \int_{z_J}^{z_I} \omega dz$. The radius of an asymptotic cylinder is given by integrating around a puncture $\oint d w = \oint_{z_i} \omega(z) dz = 2\pi i r_i$. Near the punctures we have $w = r_i \log(z-z_i)$, and so the puncture is mapped to $\tau=\pm \infty$ if the radius $r_i$ is negative/positive respectively. 

The lengths of intermediate cylinders are given by integrating along cycles at constant time $\tau$, which gives $\oint d w = \oint \omega(z) dz = 2 \pi i \r_i$. Integration over cycles with non-constant $\tau$ allows us to reconstruct the twist angles $\theta$, see \cite{Wol86} for additional details. The above procedure gives us the moduli $(\tau, \theta, \r)$ which uniquely fixes a lightcone diagram for a given Riemann surface $\Sigma$, but the reverse direction also holds in mapping a lightcone diagram to a Riemann surface\cite{Wol86}.
To summarize, given a punctured Riemann surface $\Sigma$ we can equip it with a unique Euclidean lightcone metric. Integrating over the moduli space of punctured Riemann surfaces $\mathcal{M}_{g,n}$ is thus equivalent to integrating over the moduli space of Euclidean lightcone diagrams.
\be 
\int_{\mathcal{M}_{g,n}} \ldots = \int[d \tau][d \theta][\r d \r] \ldots
\ee
In the above we have left out the integration measure to which we return to in section \ref{sec:3}.

%%%%%%%%%%%%%%%%%%%%%%%%%%%%%%%%%%%%%%%%%%%%%%%%%%%%%%%%%%%%%%%%%%%%%%%%%%%%%%%%%%%%%%%%%%%%%%%%%%%%%%%%%%
\subsection{Negative curvature lightcone diagrams} \label{sec:Negative_Curvature_Lightcone}
We are ultimately interested in JT gravity, so we want to construct the constant negative curvature analogues of lightcone diagrams. We would like to start with a Euclidean/Lorentzian lightcone metric $\hat{g}$ and turn it into a constant negative curvature geometry $g=e^{2\omega} \hat{g}$ with a Weyl factor $\omega$, with suitable singularities at the degenerate points. The reason for this is that when considering JT gravity we will find we have a constraint $R=-2$ away from the degenerate points, but there is no condition on the behavior of $\omega$ at such points. We find that the Weyl factor is not unique, and we have a large class of geometries that differ by the strength of the conical singularity at the degenerate points. The only constraint is that the resulting geometries satisfy Gauss-Bonnet.

\subsubsection*{Negative curvature Euclidean lightcone diagrams}
Since $\hat{g}$ already satisfies Gauss-Bonnet, we will choose Weyl factor to weaken the delta functions at the splittings points, and redistribute the curvature uniformly so that Gauss-Bonnet remains satisfied. Consider a Riemann surface $\Sigma$ where we prescribe conical singularities of opening angles $2\pi \alpha_I$ at points $z_I$, where the metric locally behaves as $ds^2 = |z|^{2(1-\alpha_I)} dz d \bar{z}$ near $z_I$. Does there exist a constant negative curvature metric $g$ on $\Sigma$ with the prescribed singularity structure? The answer turns out to be yes, assuming the specification of $\alpha_I$ does not violate Gauss-Bonnet
\be
\frac{1}{4 \pi} \int_{\Sigma\backslash\{z_i\}} \sqrt{g} R =  \chi(\Sigma) + \sum_I \left( 1-\alpha_I \right) < 0. 
\ee
Under the above condition, it was shown\footnote{For a review of various existence and uniqueness theorems on metrics with conical singularities see \cite{lai2016metric}.}\cite{troyanov1991prescribing,mcowen1988point} that the metric $g$ exists, is unique, and is in the conformal equivalence class of metrics on $\Sigma$. We conclude that starting from a Euclidean lightcone metric $\hat{g}$, we can find a unique $\omega$ such that $g=e^{2\omega} \hat{g}$ has constant negative curvature and any desired singularity structure consistent with Gauss-Bonnet
\be \label{eqn:euclidean_weyl}
\frac{1}{2} \sqrt{g} \left( R + 2 \right)= - 2\pi \sum_{I=1}^{2g-2+n}(1-\alpha_I) \delta^{2}(z-z_I). 
\ee
We have written the curvature in this way to emphasize that the effect of $\alpha_I$ is to take curvature from the interaction points and redistribute it uniformly across the rest of the surface $\Sigma \backslash \{z_i\}$. Note that this amounts to studying JT gravity with conical excesses, which appears to be the closest relative to Euclidean lightcone diagrams. However, the utility of lightcone diagrams is they can be easily analytically continued to singular Lorentzian geometries, which we now discuss.

\subsubsection*{Negative curvature Lorentzian lightcone diagrams}
We would like to construct negative curvature Lorentzian lightcone diagrams in a similar way by rescaling the metric $g=e^{2\omega} \hat{g}$, where $g, \hat{g}$ are almost Lorentzian geometries. As far as we are aware, not much is known about almost Lorentzian metrics and the existence of such Weyl factors. However, we would like to argue that it is plausible that there exists an appropriate $\omega$ such that the curvature of $g=e^{2\omega} \hat{g}$ takes the form 
\be \label{eqn:metric_LJT}
\frac{1}{2} \sqrt{-g} \left( R + 2 \right)=  2\pi \sum_{I=1}^{2g-2+n} \left( i+\alpha_I \right) \delta^{2}(z-z_I),
\ee
under the condition that $\alpha_I$ is chosen so that the complex Gauss-Bonnet theorem is not violated, which requires that $\alpha_I > 0$. The purpose of the $\alpha_I$ can can be seen from the form of Gauss-Bonnet for complex metrics\cite{Witten:2021nzp, Louko:1995jw}
\be \label{eqn:complex_GB}
\frac{1}{2} \int_{\Sigma} \sqrt{-g} R=-2 \pi i \chi(\Sigma).
\ee
The effect of the $\alpha_I$ is to cancel out the contribution of the bulk volume 
\be
\frac{1}{4 \pi}\int_{\Sigma\backslash \{z_I\}} \sqrt{-g} R + \sum_{I=1}^{2g-2+n} \alpha_{I} = 0,
\ee
so that the remaining terms in \eqref{eqn:metric_LJT} give the Euler characteristic. While we cannot prove that such Weyl factor exists, we think it is reasonable that it does for the following reasons. It's existence would be the natural extension of the Euclidean results of \cite{troyanov1991prescribing,mcowen1988point} which gives the unique metric \eqref{eqn:euclidean_weyl}. In both cases the role of the $\alpha_I$ is to cancel out the bulk volume term to satisfy Gauss-Bonnet. Furthermore, since the Lorentzian lightcone metric $\hat{g}$ is complex allowable, so is $e^{2 \omega} \hat{g}$, which implies that Gauss-Bonnet \eqref{eqn:complex_GB} holds for both metrics. If $e^{2 \omega} \hat{g}$ is to have constant negative curvature, it's singular points must contribute in a way to cancel out the bulk volume, otherwise Gauss-Bonnet would be violated. Finally, in appendix \ref{sec:Lorentzian_Pants} we construct the Lorentzian constant negative curvature pairs of pants with the same singularity structure as advocated above.

To summarize, we believe it is reasonable that there exists a Weyl factor that turns a Lorentzian lightcone diagram into a constant negative curvature analogue given by \eqref{eqn:metric_LJT}, with the only constraint on $\alpha_I$ being that Gauss-Bonnet \eqref{eqn:complex_GB} is satisfied. One important point is that while flat lightcone diagrams are unique, the constant negative curvature analogues are not. When integrating over the Weyl factor we must make a choice of which geometries to include, which amounts to deciding which $\alpha_I$ to include. We are not able to give a conclusive answer to this question, but we return to it in the discussion.

%%%%%%%%%%%%%%%%%%%%%%%%%%%%%%%%%%%%%%%%%%%%%%%%%%%%%%%
\section{JT gravity on lightcone diagrams} \label{sec:3} JT gravity is a two dimensional dilaton gravity theory with metric $g_{\mu \nu}$ and dilaton $\Phi$. We will start with the Euclidean JT gravity path integral with boundary conditions given by $n$ geodesic circles of given lengths $\Vec{b}=(b_1, \cdots, b_n)$. We assume the bulk geometry has genus $g$ and is connected, with the extension to the full topological expansion easily following. This amplitude can be computed by performing the path integral over all compact geometries $\Sigma$ of genus $g$ with $n$ vertex operators $V$ inserted\footnote{The other way to compute this amplitude is to perform the path integral over surfaces with $n$ boundaries with appropriate boundary conditions. However, the formalism of lightcone diagrams most easily applies to the situation where boundary conditions are specified by vertex operator insertions.}
\be \label{eqn:ZPolyakov}
Z = \int \frac{\mathcal{D}g_{\mu \nu} \mathcal{D}\Phi}{\text{Vol}} e^{-I_{\jt}[g,\Phi]} V_1 \ldots V_n,
\ee
where $\text{Vol}$ is the volume of the diffeomorphism group. The Euclidean JT action on a compact surface $\Sigma$ is given by \cite{Jackiw:1984je,Teitelboim:1983ux,Maldacena:2016upp}
\be
I_\jt[g,\Phi] = -\frac{S_0}{4\pi} \int_{\Sigma} \sqrt{g} R -\frac{1}{2} \int_\Sigma \sqrt{g} \Phi (R+2).
\ee
The first term is purely topological. Taking into account the vertex operators, the total topological contribution will be given by $e^{S_0 \chi}$ where $\chi = 2 - 2g - n$. We ignore this term from now on, although at certain points various constants will be absorbed into the definition of $S_0$. The vertex operator that inserts a geodesic boundary of length $b$ is given by \cite{Blommaert:2021fob}
\be
V = e^{-S_0}\int_\Sigma d^2 z \sqrt{g} e^{-2\pi \Phi} \cos \left( b \Phi\right).
\ee
The path integral over all two dimensional Euclidean metrics can be decomposed into an integral over the moduli space $\mathcal{M}_{g}$ of genus $g$ Riemann surfaces, and over the Weyl factor $\omega$ of the metric \cite{DHoker:1985een,HokerPhong88}
\be \label{eqn:gaugefix1}
\int \frac{\mathcal{D}g_{\mu \nu} \mathcal{D}\Phi }{\text{Vol}} = \int_{\mathcal{M}_{g}} {\mathrm d} \mu \int \mathcal{D} \omega \mathcal{D}\Phi e^{-25 S_L[\omega, \hat{g}]}.
\ee
The above path integral is performed over metrics $g = e^{2 \omega} \hat{g}$, where $\hat{g}$ is a choice of representative metric\footnote{The standard choice is to pick $\hat{g}$ such that it has constant negative curvature for $g\geq 2$.} for each point in the moduli space $\mathcal{M}_g$. In the above integral ${\mathrm d} \mu$ is the standard integration measure on the moduli space $\mathcal{M}_g$ \cite{DHoker:1985een,HokerPhong88}, since it is quite complicated we write it schematically and go into additional detail in appendix \ref{sec:measure}. The Liouville action $S_L$ will turn out to be proportional to the Euler characteristic after we integrate out the dilaton, and we absorb it into the definition of the topological term. This integral treats the moduli of the Riemann surface on a different footing from the positions of the vertex operators. Giddings and D'Hoker \cite{GidHok87} gave an argument that the integral over vertex operator positions $\int d^2 z_i \sqrt{g}$ can be absorbed into the integral over $\mathcal{M}_g$ to give an integral over the moduli space of punctured Riemann surfaces $\mathcal{M}_{g,n}$ of genus $g$ with $n$ punctures\cite{GidHok87}
\be
\int_{\mathcal{M}_{g}} {\mathrm d} \mu \prod_{i=1}^n  \left( e^{-S_0} \int d^2 z_i \sqrt{g} e^{-2\pi \Phi} \cos \left( b_i \Phi\right) \right) = \int_{\mathcal{M}_{g, n}} {\mathrm d} \mu \prod_{i=1}^n W_i.
\ee
The measure ${\mathrm d} \mu$ on the right is now for the moduli space of punctured Riemann surfaces $\mathcal{M}_{g,n}$, and various factors have been absorbed into the definition of the wavefunctions $W_i$. The integral over $\mathcal{M}_{g,n}$ automatically takes into account the integral over all distinct points $z_i$ at which the wavefunctions can be inserted, and it could have been our starting point for the Euclidean path integral\footnote{The two are equal up to overall normalization factors that can be absorbed, see \cite{GidHok87,HokerPhong88}.} \eqref{eqn:ZPolyakov}. On the right hand side we have defined the wavefunctions
\be
W_i = e^{-2\pi \Phi(z_i)} \cos \left( b_i \Phi(z_i) \right),
\ee
located at the punctures at points $z_i$ on the surface. The role that the wavefunctions $W_i$ play is that they impose boundary conditions at the punctures in the following way. Consider performing the path integral over the dilaton near a particular puncture $z_i$ with wavefunction $W_i$. When we integrate over the dilaton $\Phi$ near the puncture we get the following constraint on the behavior of the metric
\be \label{eqn:bc}
\frac{1}{2} \sqrt{g} (R+2) = (2 \pi \pm i b_i ) \delta^2 (z-z_i).
\ee
Each wavefunction comes with two branches denoted by $\pm i b$, which arise from the expansion of $\cos (b \Phi)$, and so the constraint is a linear combination of the two branches\footnote{It might at first be surprising that a geodesic boundary is given by an imaginary conical defect since a geodesic boundary does not give a delta function of curvature. This was explained in Appendix A of \cite{Blommaert:2021fob}. A geodesic boundary is given by the end of the trumpet geometry, after reaching the geodesic on the trumpet the radial coordinate can be analytically continued into the imaginary axis to reach an imaginary defect. We should thus imagine that the geometry is continued into the complex radial direction near the insertion points.}. Since each wavefunction comes with two branches, performing the dilaton integral over the entire surface gives a sum over all possible branches of the constraints.

The end result is we have re-written the path integral over two dimensional Euclidean metrics with vertex operator insertions as a path integral over the moduli space of punctured Riemann surfaces with wavefunctions inserted at the punctures
\be \label{eqn:3.8}
Z = \int_{\mathcal{M}_{g, n}} {\mathrm d}\mu \int \mathcal{D} \omega \mathcal{D} \Phi e^{-I_\jt [e^{2\omega} \hat{g}, \Phi]} W_1 \ldots W_n.
\ee
In the above integral ${\mathrm d} \mu$ is the standard measure on the moduli space $\mathcal{M}_{g,n}$ \cite{GidHok87,HokerPhong88}. The path integral measure for $\mathcal{M}_{g,n}$ is quite complicated and is written out in appendix \ref{sec:measure} alongside additional details on the moduli space.

\subsection*{Lightcone diagrams}
To perform the above integral we must choose a representative metric $\hat{g}$ for each point in $\mathcal{M}_{g,n}$, and then perform the integral over the Dilaton $\Phi$ and Weyl factor $\omega$. At this point we must make a choice to separate the $n$ wavefunctions $W_i$ into those that run to the Euclidean past or future\footnote{In bosonic string theory we would send the vertex operators defining the in state to the Euclidean past, while vertex operators that define the out state would be sent to the future.}. We must have at least one wavefunction in the future and one in the past for the Euclidean lightcone metric $\hat{g}$ to be a good gauge choice. Using the results summarized in section \ref{sec:2}, we choose the representative metric $\hat{g}$ to be given by the unique Euclidean lightcone diagram for each point in $\mathcal{M}_{g,n}$, which we write below for convenience
\be 
\frac{1}{2}\sqrt{\hat{g}} \hat{R}= - 2 \pi \sum_{I=1}^{2g-2+n} \delta^2(z-z_I), \qquad \qquad  \det \hat{g}(z_I)=0.
\ee

As explained in section \ref{sec:2}, the Euclidean lightcone metric $\hat{g}$ has a globally defined Euclidean time $\tau$. In terms of the $\tau$ coordinate on $\Sigma$, the wavefunctions $W_i$ we choose to define the in state are located at $\tau=-\infty$, while those that define the out state are at $\tau=\infty$. In lightcone diagram gauge the measure ${\mathrm d} \mu$ takes a particularly simple form\cite{GidHok87,HokerPhong88}
\be \label{eqn:gaugefixmeasure}
\int_{\mathcal{M}_{g, n}} {\mathrm d} \mu =\int[d \tau][d \theta][\r d \r] \frac{2 \pi  \operatorname{det}' (- \hat{\nabla}^2 )}{\int_{\Sigma} d^{2} z \sqrt{\hat{g}}},
\ee
up to a term proportional to the Euler characteristic that we absorbed into $S_0$. The prime on the determinant indicates that we exclude zero modes. We include more details on the derivation of the above measure in appendix \ref{sec:measure}. The determinant of the Laplacian is defined to be with respect to the Euclidean lightcone metric $\hat{g}$ on $\Sigma$, and we discuss the determinants in more detail slightly later. We end up with the following path integral 
\be \label{eqn:ZJT_Euclidean}
Z = \int_{W_i} [d \tau][d \theta][\r d \r] \frac{2 \pi  \operatorname{det}' (- \hat{\nabla}^2 )}{\int_{\Sigma} d^{2} z \sqrt{\hat{g}}} \int \mathcal{D} \omega \mathcal{D} \Phi e^{-I_\jt [e^{2\omega} \hat{g}, \Phi]}.
\ee
We have denoted the integral over lightcone diagrams as $\int_{W_i}$ to emphasize that the wavefunctions $W_i$ impose boundary conditions at the ends of the cylinders at $\tau=\pm \infty$. Note that so far all we have done is rewritten \eqref{eqn:ZPolyakov} in a particular gauge choice. This is essentially the argument for the equivalence of the Polyakov formulation of bosonic string theory and the interacting string picture \cite{Mandelstam:1973jk,Mandelstam:1985ww} on Euclidean lightcone diagrams \cite{GidHok87,Gid87}. The only new ingredients we have encountered in considering this formalism for JT gravity is that we have to integrate over the Weyl factor, and there is no canonical choice for separating the boundaries into those that are sent to the past or future. 

\subsection{Lorentzian JT path integral} \label{sec:3.1}
Since Euclidean lightcone diagrams come with a global notion of Euclidean time $\tau$, we can analytically continue $\tau \to i \tau$ to get Lorentzian lightcone diagrams with metric
\be  \label{eqn:sec3_LLC}
\frac{1}{2}\sqrt{- \hat{g}} \hat{R} = 2 \pi i \sum_{I=1}^{2g-2+n}\delta^2(z-z_I), \qquad \qquad  \det \hat{g}(z_I) =0. 
\ee
Our \emph{\textbf{definition}} for the Lorentzian JT path integral will be given by the above analytic continuation applied to \eqref{eqn:ZJT_Euclidean}, which gives us
\be
Z_{\text{L}} = \int_{W_i} [d \tau][d \theta][\r d \r] \frac{2 \pi  \operatorname{det}' (- \hat{\nabla}^2 )}{\int_{\Sigma} d^{2} z \sqrt{\hat{g}}} \int \mathcal{D} \omega \mathcal{D} \Phi e^{i I_\jt [e^{2\omega} \hat{g}, \Phi]}.
\ee
The Lorentzian JT action is now given by 
\be
I_\jt[g,\Phi] = \frac{S_0}{4\pi} \int_{\Sigma} \sqrt{-g} R +\frac{1}{2} \int_\Sigma \sqrt{-g} \Phi (R+2).
\ee
where we use the notation $g=e^{2\omega}\hat{g}$. We will ignore the first term which is topological by complex Gauss-Bonnet \eqref{eqn:complex_GB} and focus on the second term. On a lightcone diagram we use \eqref{eqn:sec3_LLC} to find that the second term is given by
\be
\frac{1}{2} \int_\Sigma \sqrt{-g} \Phi (R+2) = 2\pi i \sum_{I=1}^{2g-2+n} \Phi(z_I) + \frac{1}{2} \int_{\Sigma \backslash \{z_I\}} \sqrt{-g} \Phi \left(R+2 \right).
\ee
We see that the JT action on a lightcone diagram picks up point terms at the degenerate points where the geometry undergoes a topology changing transition. The usual JT gravity constraint of constant negative curvature $R=-2$ no longer holds at such points. If the constraint did hold on all of $\Sigma$ the path integral would be zero since a suitable Lorentzian topology changing metric does not exist. There is one issue, if we integrate over the dilaton $\Phi(z_I)$ at the interaction point the path integral will diverge. However, since there are $2g-2+n$ interaction points we can absorb this infinite divergence into the definition of $S_0$ in the topological term. 

We can now perform the path integral over the dilaton $\Phi$. Since we are in Lorentzian signature, we choose our contour for $\Phi$ to be along the real axis. Rewriting the action in terms of the Weyl factor we find
\be \label{eqn:constraint}
\int \mathcal{D} \Phi \exp\left(i \int_{\Sigma \backslash \{z_I\}} \sqrt{-\hat{g}} \Phi \left(-\hat{\nabla}^2 \omega + e^{2\omega} \right) \right) = \delta\left(-\hat{\nabla}^2 \omega + e^{2 \omega} \right) = \frac{\delta \left(\omega-\omega_0 \right)}{\det \left( -\hat{\nabla}^2 + 2 e^{2\omega_0} \right)}.
\ee
Here $\omega_0$ is a configuration that satisfies the delta function constraint. This constraint enforces that the metric $g=e^{2\omega}\hat{g}$ has constant negative curvature $R=-2$ away from the interaction points. The end result is the following path integral
\be \label{eqn:2ndtolaststep}
Z_{\text{L}} = \int_{W_i} [d \tau][d \theta][\r d \r] \frac{2 \pi  \operatorname{det}' (- \hat{\nabla}^2 )}{\int_{\Sigma} d^{2} z \sqrt{\hat{g}}} \int \mathcal{D} \omega \frac{\delta \left(\omega-\omega_0 \right)}{\det \left( -\hat{\nabla}^2 + 2 e^{2\omega_0} \right)}.
\ee
The reason we have not carried out the integral over $\omega$ is the following. As explained in section \ref{sec:Negative_Curvature_Lightcone}, there exist multiple Weyl factors that satisfy the constraint \eqref{eqn:constraint} while having different behavior at the degenerate points given by \eqref{eqn:metric_LJT}, which we restate here for convenience
\be 
\frac{1}{2} \sqrt{-g} \left( R + 2 \right)=  2\pi \sum_{I=1}^{2g-2+n} \left( i+\alpha_I \right) \delta^{2}(z-z_I). 
\ee
We must choose our contour of integration for $\omega$ to decide which configurations should be included. A configuration is specified by picking out a preferred choice of $\alpha_I$. There appear to be two natural choices. We can choose a certain $\alpha_I$ as special and the Weyl factor $\omega_0$ can be forced to localize onto such a geometry for all lightcone diagrams. The other obvious option is to integrate over all possible $\omega$ (i.e. all $\alpha_I$). The benefit of choosing a particular $\alpha_I$ as special would be that we would get the closest analogue to Euclidean JT gravity where only one $\omega$ contributes for each point in moduli space $\mathcal{M}_{g,n}$. For simplicity we will assume that the Weyl factor localizes to a single configuration, after which we are left with an integral over the moduli space $\mathcal{M}_{g,n}$ with given measure
\be \label{eqn:LJT_Final}
Z_{\rm L} = \int_{W_i} [d \tau][d \theta][\r d \r] \frac{2 \pi  \operatorname{det}' (- \hat{\nabla}^2 )}{\int_{\Sigma} d^{2} z \sqrt{\hat{g}}}  \frac{1}{\det \left( -\hat{\nabla}^2 + 2 e^{2\omega_0} \right)}.
\ee
The integration range is defined in \eqref{eqn:int_range}, and the wavefunctions $W_i$ implement boundary conditions as discussed around \eqref{eqn:bc}. The end result is that we must choose a set of boundary conditions and then evaluate the above integral. 

We now discuss the definition of the functional determinants appearing in the above integral. Even though the integral is over Lorentzian lightcone geometries, we will choose the functional determinants to be analytically continued and defined on the corresponding Euclidean lightcone geometries to make them well defined. The Laplacian $\hat{\nabla}^2$ is defined with respect to the degenerate lightcone metric \eqref{eqn:metric_ELC}. The corresponding determinant $\operatorname{det} (- \hat{\nabla}^2 )$ has been studied by \cite{PhongDets89,Sonoda:1987ra,hillairet2016spectral,kokotov2009}, and in principle it is known in terms of basic objects on the underlying punctured Riemann surface. However, the second determinant is more complicated. Using the conformal anomaly we can relate it to $\operatorname{det} (- \nabla^2 + 2 )$ where $\nabla^2$ is now defined with respect to a constant negative curvature metric $g=e^{2\omega} \hat{g}$ with conical singularities at the splitting points. There has been some recent progress on computing closely related determinants\footnote{In \cite{teo2021resolvent} various determinants were studied on punctured Riemann surfaces with conical singularities of opening angles $\frac{2\pi}{n+1}$ with $n \in \mathbb{Z}^{+}$. However, in our case we require conical excesses instead of defects so the results do not appear immediately applicable.} \cite{teo2021resolvent}, but as of now determinants on singular Riemann surfaces are not fully understood. Regardless, evaluating the above determinants will introduce some dependence on the moduli into the above integral.

\subsection*{Relation to Euclidean amplitudes}
We would like to discuss the relation of our proposed Lorentzian path integral \eqref{eqn:LJT_Final} to the standard Euclidean amplitudes. We first briefly review why the Euclidean JT path integral gives Weil-Petersson volumes, originally explained in \cite{Saad:2019lba}. For simplicity, consider a compact surface of genus $g$, the Euclidean path integral over all smooth metrics on this surface reduces to \cite{Saad:2019lba}
\be \label{eqn:Euclidean_dets}
Z = \int_{\mathcal{M}_g} {\mathrm d} {\rm (WP)} (\operatorname{det} \hat{P}_{1}^{\dagger} \hat{P}_{1})^{1 / 2} \int \mathcal{D} \omega \frac{\delta(\omega)}{\operatorname{det} (- \hat{\nabla}^2 + 2 )},
\ee
where we have gauge fixed to a smooth metric $\hat{g}$ of constant negative curvature, and as a result the Weyl factor localizes to $\omega=0$. We have written the integral in this way to compare to \eqref{eqn:2ndtolaststep}. The measure ${\mathrm d} {\rm (WP)}$ is known as the Weil-Petersson measure. In the above we have two determinants: $(\operatorname{det} \hat{P}_{1}^{\dagger} \hat{P}_{1})^{1 / 2}$ which originates from gauge fixing and is theory independent, and $\operatorname{det} (- \hat{\nabla}^2 + 2 )$ which is special to JT gravity and arises from the integral over the dilaton $\Phi$. In \cite{Saad:2019lba} it was pointed out that the ratio of these determinants is unity, up to a factor that can be absorbed into the coupling constant $S_0$. We are left with an integral over the Weil-Petersson measure, which gives us the Weil-Petersson volume
\be
Z = \int_{\mathcal{M}_g} {\mathrm d} {\rm (WP)} = V_g.
\ee
We can similarly ask whether our Lorentzian path integral \eqref{eqn:LJT_Final} localizes to the Weil-Petersson measure. Possible issues might arise both from the theory independent measure \eqref{eqn:gaugefixmeasure}, and from the theory dependent contribution \eqref{eqn:constraint}. However, choosing lightcone diagrams as the representative metric for the moduli space $\mathcal{M}_{g,n}$ is simply a gauge choice, so the theory independent measure \eqref{eqn:gaugefixmeasure} cannot depend on this choice. Indeed, it was pointed out in \cite{GidHok87} that \eqref{eqn:gaugefixmeasure} secretly contains the Weil-Petersson measure ${\mathrm d} {\rm (WP)}$ in lightcone coordinates, and that the determinant of the Laplacian $\operatorname{det} (- \hat{\nabla}^2 )$ roughly comes from the punctured Riemann surface analogue of the gauge fixing determinant $(\operatorname{det} \hat{P}_{1}^{\dagger} \hat{P}_{1})^{1 / 2}$.

The only possible issue with recovering the Euclidean amplitudes can therefore be from the determinant arising from the integral over the Weyl factor in \eqref{eqn:constraint}. However, this is precisely where we ran into an ambiguity since $\omega$ is not constrained to behave in any particular way at the degenerate points. Therefore the determinant in \eqref{eqn:constraint} should somehow depend on the choice of singularity structure at the degenerate points, and it's unclear if we get an exact cancellation as in \eqref{eqn:Euclidean_dets} for Euclidean JT. Without a better understanding of determinants on singular surfaces it is difficult to say anything more concrete. However, since the only deviation comes from a finite set of points we believe it's quite likely that the resulting amplitudes computed by \eqref{eqn:LJT_Final} share similar features with Euclidean JT gravity, we return to this point in the discussion.

%%%%%%%%%%%%%%%%%%%%%%%%%%%%%%%%%%%%%%%%%%%%%%%%%%%%%%%%%%%%%%%%%%%
\section{Discussion}
In this work we have proposed a definition for the Lorentzian JT gravity path integral that includes topology changing configurations. This is accomplished by integrating over metrics which are Lorentzian signature everywhere except at special points where the metric becomes degenerate, and where the spatial topology changing transitions occur. 

Our proposal is inspired by a formulation of bosonic string theory on singular Lorentzian worldsheets \cite{Mandelstam:1973jk,Mandelstam:1985ww,Wol86,GidHok87,Sonoda:1987ra}, and we followed similar logic to define the Lorentzian JT path integral. Using lightcone diagrams we analytically continued the Euclidean path integral to define a Lorentzian path integral over degenerate Lorentzian metrics. This analytic continuation gives a Lorentzian interpretation for the Euclidean path integral genus expansion. We end with a few comments and potential future directions.

\subsection*{Relation to Euclidean JT and degenerate points}
We found that the most serious ambiguity in the definition of our Lorentzian theory is how to properly treat the degenerate points. There appears to be no standard prescription for dealing with such points in Lorentzian signature, although see \cite{Marolf:2022ybi} for a recent discussion. In Euclidean signature the degenerate points can always be removed with a singular Weyl factor to give a smooth Euclidean metric. However, in Lorentzian signature this cannot be done since the degenerate points are crucial for the existence of an almost Lorentzian metric with spatial topology change. This gives us some freedom to choose the behavior of the metric at such points\footnote{This ambiguity does not appear when considering the bosonic string on lightcone diagrams\cite{GidHok87} since we don't integrate over the Weyl factor.} \eqref{eqn:metric_LJT}, and this choice propagates into the final path integral in \eqref{eqn:LJT_Final}.

As discussed in section \ref{sec:3.1}, to understand the relation between Lorentzian and Euclidean amplitudes we must understand how the ambiguity at the singular points modifies the path integral measure. Indeed, in both the Lorentzian and Euclidean cases the domain of integration is given by the moduli space of punctured Riemann surfaces $\mathcal{M}_{g,n}$, so the only difference can arise from the respective measures on this space. Since the determinants on singular Riemann surfaces appearing in \eqref{eqn:LJT_Final} are not fully understood it is difficult to give a conclusive answer. However, it is likely that this ambiguity modifies the path integral measure, and so we don't expect the Lorentzian amplitudes to exactly match the Euclidean amplitudes.

Regardless, we expect the Lorentzian amplitudes to have qualitatively similar behavior to Euclidean JT gravity amplitudes for reasons of random matrix universality. In \cite{Blommaert:2022lbh} it was shown that wormhole contributions in a wide range of dilaton-gravity models, with integration measures deviating from pure JT gravity, have universal behavior independent of the specific details of the theory. Indeed, we do not expect the specific details of the integration measure to seriously modify universal behavior, such as the existence of a page curve\cite{Almheiri:2019qdq,Penington:2019kki} or the spectral form factor \cite{Saad:2018bqo,Saad:2019lba,Blommaert:2022lbh}.

\subsection*{Towards non-perturbative Lorentzian physics}
Non-perturbative Euclidean wormholes have played an important role in a wide variety of recent calculations, see \cite{Almheiri:2019qdq,Penington:2019kki,Saad:2018bqo,Saad:2019lba,Blommaert:2022lbh,Saad:2019pqd,Iliesiu:2021ari,Stanford:2022fdt} for some selected examples. It would be interesting to understand the Lorentzian origin of these calculations using the Lorentzian path integral developed in this work. As an example, it would be interesting to understand the Replica wormhole computations \cite{Almheiri:2019qdq,Penington:2019kki} or the swap entropy \cite{Marolf:2020rpm} using topology changing Lorentzian wormholes, see also \cite{Colin-Ellerin:2020mva,Colin-Ellerin:2021jev}. Additionally, it would be interesting to understand how the inclusion of topology changing Lorentzian metrics modifies the structure of the JT gravity Hilbert space\cite{Harlow:2018tqv}, and whether there is a canonical quantization interpretation for Lorentzian topology change.

To address the above questions it will be necessary to introduce asymptotically AdS boundaries, which amounts to studying lightcone diagrams with boundaries. These diagrams were studied in the context of open strings\cite{Mandelstam:1973jk,Mandelstam:1985ww}, but the proof of Giddings and Wolpert \cite{Wol86} was never extended to them. Thus we cannot claim that these diagrams give a single cover of the desired moduli space, but it seems quite likely that they do. Regardless, we can directly define the Lorentzian JT theory with asymptotic AdS boundaries to live on such lightcone diagrams, but for now we leave an examination of the above questions to future work. 

\subsection*{Simpler representation of moduli space}
One of the features of lightcone diagrams is that they provide a particularly simple representation for the integral over the moduli space of punctured Riemann surfaces $\mathcal{M}_{g,n}$. The standard procedure is to introduce Fenchel-Nielsen coordinates on the moduli space in terms of which the the domain of integration is incredibly complicated, see \cite{Saad:2019lba} for a discussion. However, lightcone diagrams have a very simple region of integration \eqref{eqn:int_range}
\be
\int_{\mathcal{M}_{g,n}} \ldots = \frac{1}{S}  \prod_{i,j,k} \int_{0}^{\infty} d\tau_i \int_{0}^{2\pi} d \theta_j  \int_0^{r_{\rm max}} \r_k d \r_k \ldots
\ee
while the integration measure can be quite complicated, see \eqref{eqn:LJT_Final}. It would be interesting to understand the integration measure in \eqref{eqn:LJT_Final} in terms of the lightcone coordinates on the moduli space. It might be possible that in corners of moduli space the measure takes a simplified form in terms of these coordinates. More generally, it would be interesting to understand if the ratio of determinants in \eqref{eqn:LJT_Final} simplifies as in the case of Euclidean JT gravity.

\subsection*{Acknowledgments}
We thank Andreas Blommaert, Ven Chandrasekaran, Steve Giddings, Bob Knighton, Adam Levine, and Henry Maxfield for useful discussions. We would especially like to thank Don Marolf for numerous discussions on related topics. MU is supported in part by the NSF Graduate Research Fellowship Program under grant DGE1752814, by the Berkeley Center for Theoretical Physics, by the DOE under award DE-SC0019380 and under the contract DE-AC02-05CH11231, by NSF grant PHY1820912, by the Heising-Simons Foundation, the Simons Foundation, and National Science Foundation Grant No. NSF PHY-1748958.

%%%%%%%%%%%%%%%%%%%%%%%%%%%%%%%%%%%%%%%%%%%%%%%%%%%%%%%%%%%%%%%%%%%
\appendix
\section{Lorentzian pair of pants} \label{sec:Lorentzian_Pants}
\begin{figure}
\includegraphics[width=8cm]{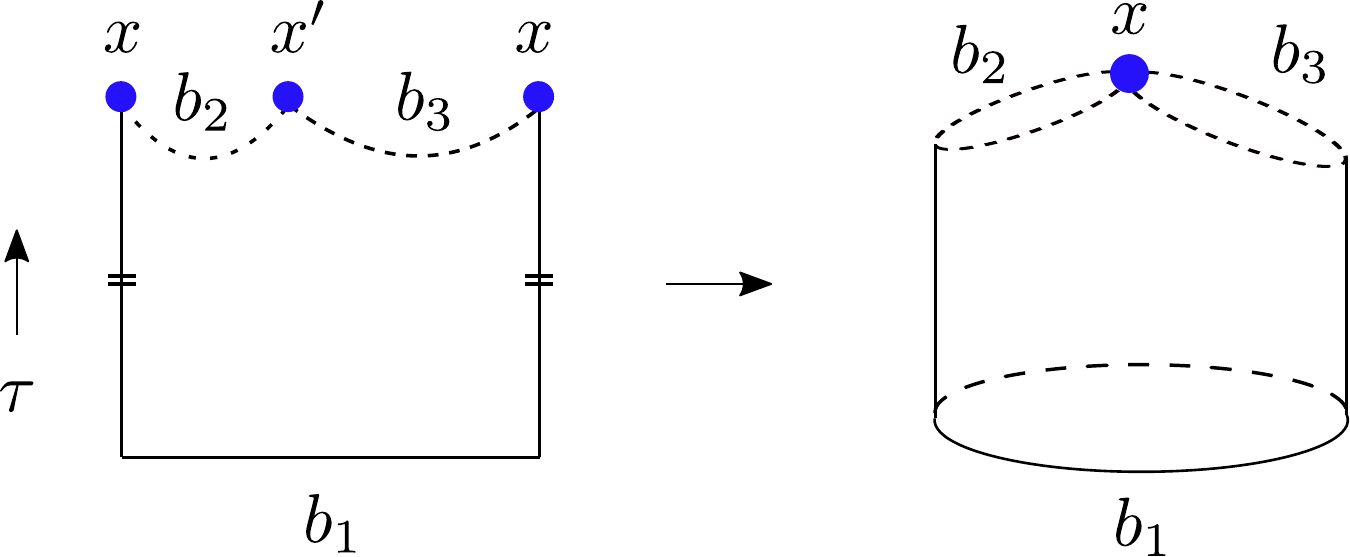}
\centering
\caption{Building the Lorentzian constant negative curvature pair of pants with geodesic boundaries by identifying points $x,x'$. The dashed lines are geodesics that form a figure eight with a singular point between them. The geometry on the right is smooth everywhere except at the point $x\sim x'$.}
\label{fig:Building_LPants}
\end{figure}

In this section we explain how to construct the constant negative curvature Lorentzian pair of pants\footnote{We thank Don Marolf for suggesting this construction.}. This can be accomplished with an identification similar to the one used for the flat pants in figure \ref{fig:wcoords}. We start with the Lorentzian $R=-2$ cylinder which is the analytic continuation of the double trumpet
\begin{equation} \label{eqn:Edoubletrumpet}
d s^{2}=\frac{- d \t^{2}+d \sigma^{2}}{\cosh^{2} \t}, \qquad \t \in\left(-\infty, \infty\right), \qquad \sigma \in[0, b_1].
\end{equation}
From the identification $\sigma \sim \sigma + b_1$ we have a spatial geodesic of length $b_1$ at $\tau=0$. We now pick two points $x,x'$ at constant $\tau$ but at different values of $\sigma$, and we connect these points with two geodesics traversing both sides of the cylinder, see figure \ref{fig:Building_LPants}. The geodesics will have lengths $b_2, b_3$ such that $b_1 \geq b_2 + b_3$. Solving for the geodesic lengths in terms of $x,x'$, it's possible to show that we can find $x,x'$ to give any geodesic lengths $b_2, b_3$ satisfying this constraint. We now identify the points $x\sim x'$ and discard the portion of the geometry above the geodesics to get a Lorentzian pair of pants with a singular point at $x\sim x'$.

When the geodesics meet at $x,x'$ there is a jump in the tangent vector, so there is a delta function in the Extrinsic curvature of the boundary curve at those points. This is taken into account in the Gauss-Bonnet theorem through the jump angles $\alpha_i$, of which there are two in the above construction
\be
\frac{1}{2}\int \sqrt{g} R + \sum_i \alpha_i = 2 \pi \chi.
\ee
We'll absorb the jump angles into a delta function contribution to the Ricci scalar at the point $x\sim x'$. 
The above is the Euclidean version of the Gauss-Bonnet theorem, the analytically continued version for almost Lorentzian metrics is given by \eqref{eqn:L_GaussBonnet} \cite{Louko:1995jw,Witten:2021nzp}
\be
\frac{1}{2} \int \sqrt{-g} R = -2 \pi i \chi.
\ee
and we see that the effect of the jump angles is to cancel out the bulk volume contribution as explained in section \ref{sec:Negative_Curvature_Lightcone}. The scalar curvature of the pair of pants is then given by
\be
\frac{1}{2} \sqrt{-g} \left( R + 2 \right)=  \left( 2\pi i+\alpha \right) \delta^{2}(x). 
\ee
where we have relabeled the sum of jump angles as just $\alpha$. One way to see that the imaginary delta function contribution appears after the identification of $x\sim x'$ is to note that without the Weyl factor in \eqref{eqn:Edoubletrumpet} this is the construction of the flat Lorentzian pants, where this contribution can be explicitly evaluated\cite{Louko:1995jw}. We can also attach pant legs to the boundaries $b_{2}$ or $b_3$ as follows. Consider a Lorentzian cylinder with two geodesic boundaries where one boundary has a corner so that the geometry exists. We can glue the boundary with the corner to the pants at $b_{2}$ or $b_3$ making sure the corner coincides with the point $x\sim x'$. This will give additional contributions to the $\alpha$ term, and gives Lorentzian pants with any desired geodesic boundary lengths.

\section{Details on punctured Riemann surfaces and measures} \label{sec:measure}
In this section we include some additional details on punctured Riemann surfaces and on the integration measure. A punctured surface $\Sigma$ is defined by taking a compact surface $\overline{\Sigma}$ of genus $g$ and removing $n$ distinct points $\Sigma = \overline{\Sigma} \backslash \{z_1, \ldots, z_n\}$, after which $\Sigma$ is no longer compact and its Euler characteristic is given by $\chi(\Sigma) = 2-2g-n$. When performing an integral over a punctured surface the punctures do not contribute to the integral. The moduli space of punctured Riemann surfaces can be constructed by considering equivalence classes of singular Euclidean metrics on $\Sigma$ \cite{Sonoda:1987ra}. We consider metrics that behave as
\be
ds^2 = \frac{c_i}{|z-z_i|^2} dz d\bar{z} + \ldots
\ee
near the punctures at $z_i$, where $c_i$ is some constant. We consider all such metrics related by smooth Weyl transformations to be in the same conformal equivalence class. The set of all such equivalence classes gives us the moduli space of punctured Riemann surfaces $\mathcal{M}_{g,n}$. In the literature it is also common to include additional singular metrics in the conformal equivalence class\cite{Wol86,HokerPhong88,PhongDets89}, such as the lightcone diagram metrics described in section \ref{sec:2}. These metrics are related to other metrics in the equivalence class through Weyl transformations that are singular at isolated points, see \cite{Mandelstam:1985ww} for an explicit example at genus zero. 

The path integral over metrics on a genus $g$ surface with $n$ punctures was analyzed in \cite{GidHok87} and is given by
\be
\int \frac{\mathcal{D}g }{\text{Vol}} = \int_{\mathcal{M}_{g, n}}[d m]  \frac{\operatorname{det}\left\langle\mu_{\alpha}, \phi_{\beta}\right\rangle}{\operatorname{det}\left(\phi_{\alpha}, \phi_{\beta}\right)^{\frac{1}{2}}} (\operatorname{det} \hat{P}_{1}^{\dagger} \hat{P}_{1})^{1 / 2} \int \mathcal{D} \omega e^{-26 S_L[\omega, \hat{g}]} .
\ee
Where on the left side $\text{Vol}$ is the volume of the diffeomorphism group of the punctured surface, and we have gauge fixed to a metric $\hat{g}$ on the right. Compare this to equation \eqref{eqn:3.8} in the main text\footnote{Note that in \eqref{eqn:gaugefix1} we are integrating over $\mathcal{M}_g$ instead of punctured surfaces $\mathcal{M}_{g,n}$. The integration measure slightly differs between the two cases, see section 3 of \cite{GidHok87}.}. In the above integral $m$ is a coordinate on the moduli space and $(\operatorname{det} \hat{P}_{1}^{\dagger} \hat{P}_{1})^{1 / 2}$ comes from gauge fixing to the representative metric $\hat{g}$, and in total the above is the integration measure for the moduli space $\mathcal{M}_{g,n}$. For a detailed explanation of the various determinants in the above measure see \cite{DHoker:1985een,HokerPhong88,GidHok87}. In the main text we introduced the shorthand notation ${\mathrm d} \mu$ for the integration measure for simplicity, we now define it in terms of the above objects for easier comparison to the literature\cite{GidHok87,DHoker:1985een,HokerPhong88}
\be
\int_{\mathcal{M}_{g, n}} {\mathrm d} \mu \equiv\int_{\mathcal{M}_{g, n}}[d m]  \frac{\operatorname{det}\left\langle\mu_{\alpha}, \phi_{\beta}\right\rangle}{\operatorname{det}\left(\phi_{\alpha}, \phi_{\beta}\right)^{\frac{1}{2}}} (\operatorname{det} \hat{P}_{1}^{\dagger} \hat{P}_{1})^{1 / 2}.
\ee

Gauge fixing to the lightcone metric $\hat{g}$, the various determinants in the above integral were worked out in \cite{GidHok87} (see equations (2.22) and (4.7-4.8) in \cite{GidHok87}, see also section V.G in \cite{HokerPhong88}). The product of determinants significantly simplifies, and we find the final result
\be
\int \frac{\mathcal{D}g }{\text{Vol}} = \int [d \tau][d \theta][\r d \r] \frac{2 \pi  \operatorname{det}' (- \hat{\nabla}^2 )}{\int_{\Sigma} d^{2} z \sqrt{\hat{g}}} \int \mathcal{D} \omega e^{-26 S_L[\omega, \hat{g}]}.
\ee
Compare this to equations \eqref{eqn:gaugefixmeasure} and \eqref{eqn:ZJT_Euclidean}. The prime on the determinant indicates that we remove zero modes.

%%%%%%%%%%%%%%%%%%%%%%%%%%%%%%%%%%%%%%%%%%%%%%%%%%%%%%%%%%%%

\bibliographystyle{utphys}
\bibliography{Refs.bib}
\end{document}